\documentclass[twocolumn,aps,showpacs]{revtex4-1}

\usepackage{amsmath}
\usepackage{amssymb}
\usepackage{graphicx}
\usepackage{hyperref}
\usepackage[usenames,dvipsnames]{color}

\begin{document}
\title{Annihilation of vortex dipoles in an Oblate Bose-Einstein Condensate}

\author{Shashi Prabhakar,$^1$ R. P. Singh,$^1$ S. Gautam,$^2$ and D. Angom$^1$}
\affiliation{$^1$Physical Research Laboratory, Navrangpura, Ahmedabad. 380009. India}
\affiliation{$^2$Indian Institute of Science, Mathikere, Bangalore. 560012. India}

\begin{abstract}
We theoretically explore the annihilation of vortex dipoles, generated when an 
obstacle moves through an oblate Bose-Einstein condensate, and examine the
energetics of the annihilation event. We show that the gray soliton, which 
results from the vortex dipole annihilation, is lower in energy than the 
vortex dipole. We also investigate the annihilation events numerically and 
observe that the annihilation occurs only when the vortex dipole overtakes the 
obstacle and comes closer than the coherence length. Furthermore, we find that 
the noise reduces the probability of annihilation events. This may explain the 
lack of annihilation events in experimental realizations.
\end{abstract}

\pacs{03.75.Lm, 03.75.Kk, 67.85.De}
\maketitle

\section{\label{sec:intr}Introduction}

One of the important developments in recent experiments on atomic Bose-Einstein 
condensates (BECs) is the creation of vortices and the study of their dynamics 
\cite{Anglin2002, PRL.83.2498}. Equally important is the recent experimental 
observation of a vortex dipole, which consists of a vortex-antivortex pair, 
when an obstacle moves through a BEC \cite{PRL.104.160401} and observation of 
vortex dipoles produced through phase imprinting \cite{F03092010, 
PRA.64.043601}. In superfluids, the vortices carry quantized angular momenta 
and are the topological defects, which often serve as the conclusive evidence 
of superfluidity. In a vortex dipole, vortices of opposite circulation cancel 
each other's angular momentum and thus carry only linear momentum. This is the 
cause of several exotic phenomena like leap frogging, snake instability 
\cite{PRA.65.043612}, orbital motion \cite{PRA.84}, trapping \cite{PRX.1}, 
and others. The effects of vortices are widespread in classical fluid flow 
\cite{CJ:392656} and optical manipulation \cite{Grier2003}. A good description 
of vortices in superfluids is given in Ref. \cite{pethick} and review articles 
\cite{rmp.81.647, rmp.59.87}. More detailed discussion of vortices is given in 
Ref. \cite{okulov}.

Among the important phenomena associated with the Bose-Einstein condensate 
(BEC), the creation, dynamics, and annihilation of vortex dipoles carry useful
information associated with the system. Several methods have been suggested to 
nucleate vortices and recently, nucleation of the vortices have been observed 
experimentally by passing a Gaussian obstacle through the BEC with a speed 
greater than some critical speed \cite{PRL.104.160401}. The trajectories of 
these vortex dipoles are ring-structured as described in Refs. 
\cite{PRA.61.013604, PRA.77.053610}. The annihilation of vortices in the 
BEC has been mentioned in a number of theoretical studies \cite{JLTP.146.31, 
PRA.84.023637, PRB.84.020506}. However, there is a lack of extensive study on 
this topic. Moreover, the questions related to thermodynamic stability, 
resulting state after annihilation and its dynamics has not been examined. The 
study of vortex dipole annihilation will shed light on the process that leads 
to minimum separation between vortex-antivortex and conditions for annihilation 
along with other phenomena arising from the dynamics of vortex dipoles.

In this work, we present analytical as well as numerical results related to 
vortex dipole annihilation for an oblate BEC at zero temperature. The results 
are obtained using Gross-Pitaevskii (GP) equation. In Section \ref{sec:theory} 
of this article, we provide a brief description of the two-dimensional (2D) GP 
equation and vortex dipole solutions. Condensate with diametric vortex dipole 
and gray soltion are studied and this is described in section 
\ref{sec:analytical}. Section \ref{sec:analytical} contains studies done in 
the strong as well as weak interacting systems. Annihilation of vortex dipoles 
is analysed from the energies obtained from the analytical calculations. The 
numerical results, confirming the analytic results, are discussed in Section 
\ref{sec:result}, and we then conclude.


\section{\label{sec:theory}Superfluid vortex dipole and its generation}

In the mean-field approximation, the dynamics of a dilute BEC is very well 
described by the GP equation 
\begin{equation}\label{hamiltonian}
   i\hbar \partial_{t}\Psi(\textbf{r},t)=[{\cal H}+U|\Psi
   (\textbf{r},t)|^{2}] \Psi(\textbf{r},t),
\end{equation}
where ${\cal H}$, $U$ and $\Psi$ are the single-particle Hamiltonian,
interaction strength and order parameter of the condensate respectively. 
The order parameter, $\Psi$, is normalized to $N$, the total number of atoms 
in the condensate. In the present case, the single-particle Hamiltonian 
${\cal H}$ consists of the kinetic-energy operator, an axis-symmetric harmonic 
trapping potential, and a gaussian obstacle potential, that is,
\begin{equation}\label{gpeqn}
   {\cal H}=-\frac{\hbar^{2}}{2m}\nabla^2+\frac{m\omega^2}{2}(x^2+
       \alpha^2y^2+\beta^2z^2)+V_{\rm obs}(x,y,t),
\end{equation}
where $\alpha$ and $\beta$ are the anisotropies along $y$ and $z$ axis 
respectively, and $V_{\rm obs}(x,y,t)$ is the repulsive Gaussian obstacle 
potential. Experimentally, a blue-detuned laser beam is used to generate
the $V_{\rm obs}(x,y,t)$ and it can be written as
\begin{equation}
   \label{vobs}
   V_{\rm obs}(x,y,t)=V_{0}(t)\exp\left[-2\frac{(x-vt)^{2}+y^{2}}
       {w_{0}^{2}} \right],
\end{equation}
where $V_{0}(t)$ is the potential at the center of the Gaussian obstacle at 
time $t$, $v$ is the velocity of the obstacle along $x$-axis, and $w_0$ is the
radius of repulsive obstacle potential. In the present work, we consider the 
motion of obstacle along $x$-axis only. Defining the oscillator length of the 
trapping potential as $a_{\rm osc}=\sqrt{\hbar/(m \omega)}$, and considering 
$\hbar\omega$ as the unit of energy, we can then rewrite the equations in 
dimensionless form with transformations $\tilde{\mathbf{r}}= \mathbf{r}/ 
a_{\rm osc}$, $\tilde{t}=t \omega$, and the transformed order parameter 
assumes the form
\begin{equation}
   \phi(\tilde{\mathbf{r}},\tilde{t})=\sqrt{\frac{a_{\rm osc}^3}{N}}
       \Psi(\mathbf{r},t).
\end{equation}
For the sake of notational simplicity, hereafter we denote the scaled 
quantities without tilde in the rest of the manuscript. In a pancake-shaped 
trap $\alpha=1$ and $\beta\gg 1$ and the order parameter can then be written 
as
\begin{equation}
   \phi(\mathbf{r},t)=\psi(x,y,t)\zeta(z)\exp(-i\beta t/2),
\end{equation}
where $\zeta(z)=(\beta/(2\pi))^{1/4}\exp(-\beta z^2/4)$. The Eq. 
(\ref{hamiltonian}) is then reduced to the two dimensional form
\begin{eqnarray}\label{gp_2d}
   \left[-\frac{1}{2}\left(\frac{\partial^2}{\partial x^2}
       +\frac{\partial^2}{\partial y^2}\right)+\frac{x^2+y^2}{2}
       +\frac{V_{\rm obs}(x,y,t)}{\hbar\omega} \right. \nonumber \\
   \left.+u|\psi({\mathbf r},t)|^2-i\frac{\partial}{\partial t}\right]
       \psi ({\mathbf r},t) = 0,
\end{eqnarray}
where $u=2aN\sqrt{2\pi\beta}/a_{\rm osc}$, with $a$ as the $s$-wave 
scattering length, is the modified interaction strength. In the present work, 
we consider condensate consisting of $^{87}$Rb atoms in $F=1$, $m_F=-1$ state 
with $s=99a_0$ \cite{prl88}. We have neglected a constant term corresponding 
to the energy along axial direction as it only shifts the energies and 
chemical potentials by a constant without affecting the dynamics. We solve
this equation numerically using the Crank-Nicholson 
method \cite{Muruganandam20091888}.

There are several theoretical and experimental proposals to generate vortices 
in non-rotating traps. These include stirring of the condensate using 
blue-detuned laser or several laser beams \cite{PRL.104.160401, PRA.61.013604}, 
adiabatic passage \cite{PRL.80.2972}, Raman transitions in binary condensate
systems \cite{PRL.79.4728}, laser beam vortex guiding \cite{s003400000337}, 
and phase imprinting \cite{PRA.64.043601}. Among these methods, the easiest 
one to nucleate vortex dipoles is by stirring a BEC with a blue-detuned laser 
beam. When the velocity of the laser beam exceeds a critical velocity, 
vortex-antivortex pairs are released from the localized dip in the number 
density created due to the laser beam. These vortex dipoles then move through 
the BEC and exhibit various interesting dynamics \cite{F03092010, 
PRA.61.013604, PRA.83.011603}. The critical velocity depends on the number 
density, width and intensity of the laser beam, and the frequency of the 
trapping potential. This nucleation process exhibits a high degree of coherence 
and stability, allowing us to map out the annihilation of the dipoles. In an 
axis-symmetric trap, a vortex dipole is a metastable state of superfluid flow 
with long lifetime.


\section{\label{sec:analytical}Condensates with vortex dipole or gray 
        soliton}

To analyse the vortex dipole annihilation, we consider a model system where the 
vortex-antivortex dipole pair and gray soliton, which may be generated when
annihilation of vortex dipole occures, are static. However, we vary the 
distance of separation and examine the energy of the total system. The present 
system can be studied under two regims: strongly interactly system, and weakly 
interacting system. The strongly interacting system is studied considering 
$\phi$ with Thomas-Fermi (TF) approximation and the weakly interacting system 
is studied considering the Gaussian form of $\phi$.


\subsection{\label{subsec:TF}Strongly interacting system with TF 
approximation}

For the $Na/a_{\rm osc}\gg1$ case, we use TF approximation to determine the 
steady state density profile and energy of the condensate. To begin with, we 
consider a condensate with vortex dipole and later, with gray soliton.


\subsubsection{\label{subsubsec:TFvd}Diametric vortex dipole}
\begin{figure}[h]
\begin{center}
   \includegraphics[width=7.5cm]{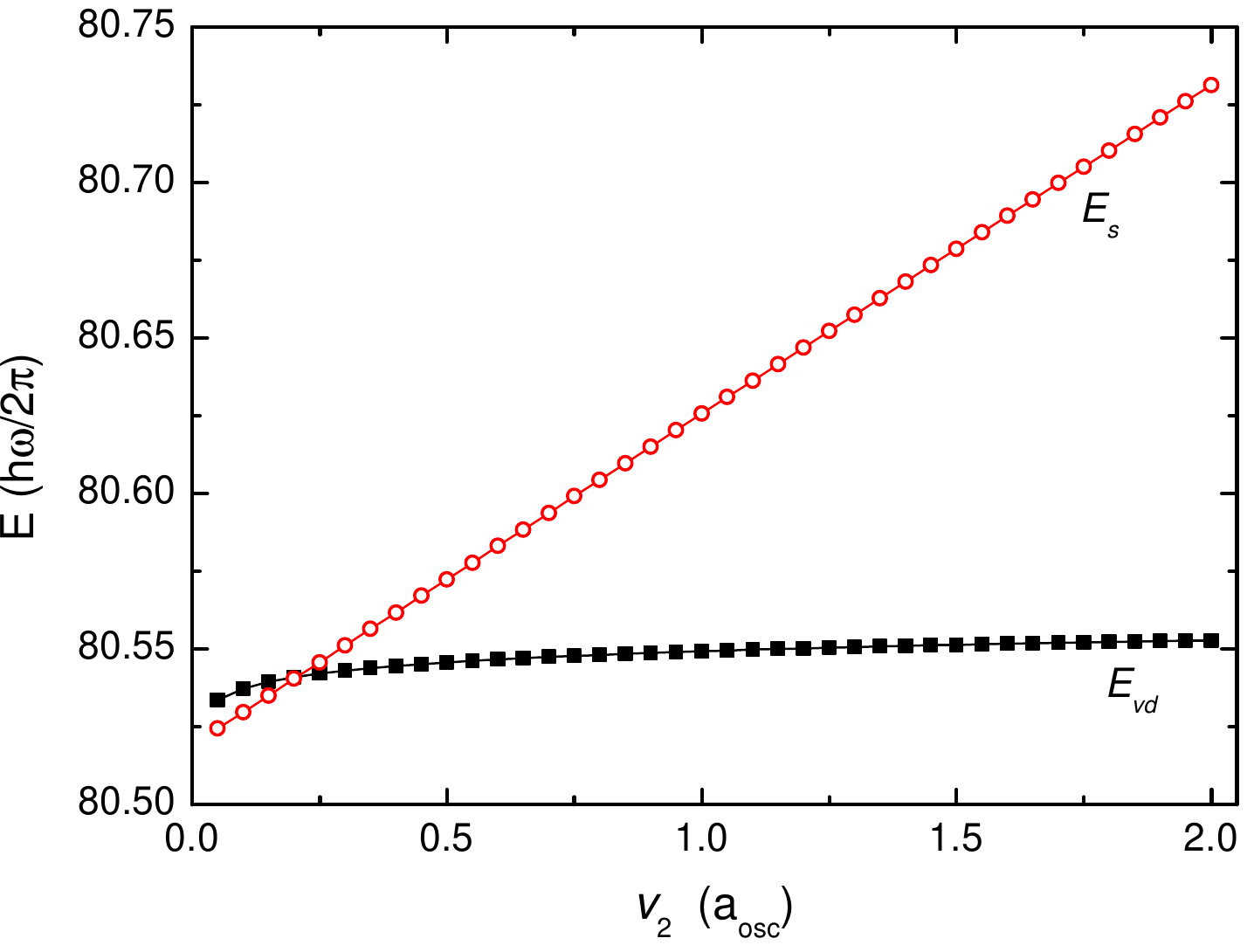}
   \caption{Comparing the energy of vortex dipole and band soliton under TF 
     approximations. The crossover in energy can be seen through ansartz 
     choosen and the analytical expressions obtained.}
\label{theoretical_pdf}
\end{center}
\end{figure}

We consider a condensate consisting of $N$ atoms in a purely harmonic potential
\begin{equation}
  V(x,y) = \frac{x^2+y^2}{2}.
\end{equation}
Consider that the condensate has a vortex dipole, consisting of a vortex and an 
antivortex located at $(0,v_2)$ and $(0,-v_2)$, respectively. The cores of the 
vortex and antivortes can be approximated as cicular regions centered around  
$(0,v_2)$ and $(0,-v_2)$ and with radii equal to the coherence lenght $\xi$. At 
the cores, we consider the density to be equal to zero. Hence, we use the TF 
approximation and adopt the following piecewise ansatz for the density of the 
condensate
\begin{eqnarray}
  n(x,y) 
  &=& \left \{ 
     \begin{aligned}
        & 0 \!&&\text{ for }  x^2+y^2 > R^2\\
        & 0 \!&&\text{ for } [x^2+(y\pm v_2)^2] \leqslant \xi^2\\
        & \left[\frac{\mu - V(x,y)}{u}\right] 
              && \text{ for } \left \{ 
                   \begin{aligned}
                     & x^2+y^2 \leqslant R^2~\&\\
                     &~[x^2+(y\pm v_2)^2] > \xi^2,
                  \end{aligned} \right .
     \end{aligned} \right .  
\end{eqnarray}
where $R = \sqrt{2\mu}$ is the spatial extent of the condensate in TF 
approximation, and $\xi = 1/R$ is the coherence length at the center of the 
trap. Normalizing this ansatz yields
\begin{equation}
  \frac{\pi \left(2-4 R^4+R^8+4 R^2 v_2^2\right)}{4 R^4 u} = 1.
  \label{normalization1}
\end{equation}
This equation defines the radius of the condensate. The TF ansatz can be used 
to calculate the total potential energy arising from the regions outside the 
cores of the vortices and is given as
\begin{equation}
  E_0 = \frac{\pi  \left[1-3 R^8+R^{12}+3 R^2 v_2^2 \left(2+R^2v_2^2\right)
    \right]}{12 R^6 u}.
\end{equation}
The main energy contribution from the vortex dipole is the kinetic energy due 
the velocity field associated with it. This energy can be approximated 
as \cite{Zhou}
\begin{equation}
  E_{\rm KE} = \frac{R^2}{u} \text{Log}\left(\frac{2v_2}{\xi}\right).
\end{equation}
This relation is valid when $\xi \ll v_2 \ll R$ and  in the present work, 
$\xi\sim0.06$, $R=15.5~a_{\rm osc}$. In order to estimate the energy 
contributions from the cores of the vortices we approximate the density within 
the cores as
\begin{equation}
  n(x,y) 
   =  \left \{ 
     \begin{aligned}
         \frac{2n_0[x^2+(y-v_2)^2]}{x^2+(y-v_2)^2+\xi^2} &
                  \text{ for } [x^2+(y- v_2)^2] < \xi^2      \\
         \frac{2n_0[x^2+(y+v_2)^2]}{x^2+(y+v_2)^2+\xi^2} &
                  \text{ for } [x^2+(y+ v_2)^2] < \xi^2,
     \end{aligned} \right.
  \label{ansatz-core-region} 
\end{equation}
where $n_0$ is the average TF density on the circle $x^2+(y\pm v_2)^2 = \xi^2$.
Assuming that the normalization is still defined by equation 
Eq.~(\ref{normalization1}), Eq.~(\ref{ansatz-core-region}) can be used to 
calculate energy contribution from the core region. The energy within the 
consist of
\begin{eqnarray}
  E^{\rm q}_{\rm c}   & = &  \frac{6 \pi  n_0}{8},\\
  E^{\rm tr}_{\rm c}  & = &  \pi  \xi ^4 (\text{Log}[4] - 1) n_0, \\
  E^{\rm int}_{\rm c} & = &  2\pi  u \xi ^2 (3-\text{Log}[16]) n_0^2,
\end{eqnarray}
where, $E^{\rm q}_{\rm c}$, $E^{\rm tr}_{\rm c}$  and $E^{\rm int}_{\rm c}$ 
are the energies arising from the quantum pressure, trapping potential and 
interaction within the core region, respectively. Thus, the total energy of 
the condensate with a vortex dipole is
\begin{equation}
  E_{\rm vd}  = E_0 + E_{\rm KE} + E^{\rm q}_{\rm c} + E^{\rm tr}_{\rm c} 
                + E^{\rm int}_{\rm c}.
  \label{E_vd}
\end{equation}
The variation of $E_{\rm vd}$ as a function of $v_2$ is plotted in Fig. 
\ref{theoretical_pdf}.

\subsubsection{\label{subsubsec:TFgs}Gray soliton}
For gray soliton extending from (0, $-v_2$) to (0, $v_2$) along $y$-axis, we 
use the following piecewise ansatz in the TF approximation
\begin{equation}
 n(x,y) 
 = \left \{ 
     \begin{aligned}
         &0 && \text{ for } x^2+y^2 > R^2,    \\
         &\left[\frac{\mu - V(x,y)}{u}\right] && \text{ for } \left \{
                \begin{aligned}  x^2+y^2 \leq R^2, \\
                                  |x|>\xi, \\
                                  |y|>v_2,
                \end{aligned} \right .\\
         &\left[\frac{\mu - V(x,y)}{u}\right] \frac{2 x^2}{x^2+\xi^2} 
          && \text{ for }  |x|\le \xi~ \& ~|y|\le v_2.
     \end{aligned} \right.
  \label{soliton-ansatz}
\end{equation}
And, the normalization condition leads to following constraint on the radius 
of the condensate
\begin{eqnarray}
  \frac{1}{12 R^3 u}& & \left [  3 \pi R^7+4 v_2 \left(10+6 R^4-3 \pi 
       \left(1+R^4\right)\right.  \right. \nonumber \\ 
    && +\left. \left . (-2+\pi) R^2 v_2^2\right ) \right ] = 1.
\end{eqnarray}
For the gray soliton, other than the quantum pressure, there is no need to 
separate out the energy associated with the trapping and interaction potential
within the soliton. So, the total energy of the system is
\begin{equation}
  E_{\rm s}  = E_0 + E^{\rm q}_{\rm c},
  \label{E_s}
\end{equation}
where, $E_0$ is the potential energy associated with the system and 
$E^{\rm q}_{\rm c} $ is the energy arising from the quantum pressure. 
These are given as 
\begin{eqnarray}
  E_0 & = & \int\int \left[V(x,y)n(x,y) + \frac{u}{2} n(x,y)^2\right] 
      \mathrm{d}x \mathrm{d}y,\nonumber\\
  E^{\rm q}_{\rm c}  & = & \frac{1}{2}\int_{-\xi}^{\xi}\left[\int_{-v_2}^{v_2} 
      |\nabla_{xy} \sqrt{n(x,y)}|^2 \mathrm{d}y\right]\mathrm{d}x.
   \label{E_0-and-QP}
\end{eqnarray}
From the expression of the $ n(x, y)$ in Eq. (\ref{soliton-ansatz}), we get
\begin{eqnarray}
  E_0 & = &\frac{1}{180 R^5 u}\left \{15 \pi  R^{11}+3 \left[236-75 
                \pi \right.\right.\nonumber\\
      &  & \left.+20 (19-6 \pi ) R^4+15 (8-3 \pi ) R^8\right] v_2 
                \nonumber\\
      &  & \left.+10 R^2 \left[-28+9 \pi +6 (-3 + \pi ) R^4 \right] 
                v_2^3\right.\nonumber\\
      &  & \left.-9 (-4+\pi ) R^4 v_2^5\right\} \\
  E^{\rm q}_{\rm c} &= &-\frac{(8+3 \pi ) v_2 \left(-3 R^2
          +3 \xi^2+v_2^2\right)}{48 u \xi}.
\end{eqnarray}
Interestingly, the $E^{\rm q}_{\rm c} $ has a $1/\xi$ dependence, which is to
be expected as smaller $\xi$ implies larger density variation and 
translates to higher quantum pressure. 

\begin{figure}[h]
\begin{center}
   \includegraphics[width=8.3cm]{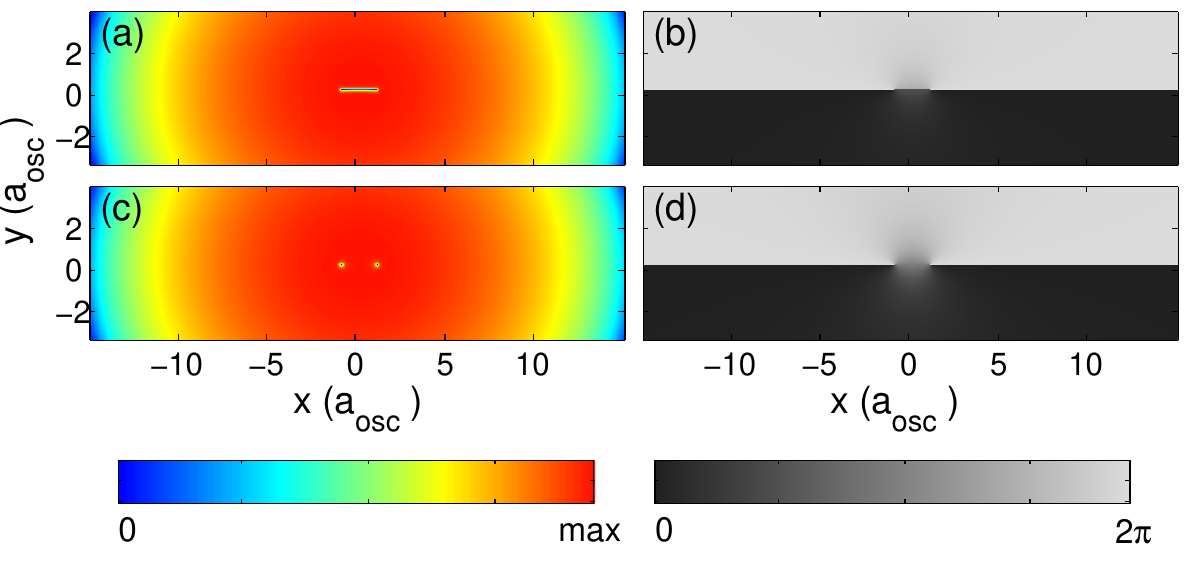}
   \caption{Band soliton (top) and vortex dipole (bottom) with density profile 
     (left) and phase profile (right) obtained numerically.}
     \label{soliton_dipole}
\end{center}
\end{figure}

For illustration, the vortex dipole and gray soliton inside the condensate is 
shown in Fig. \ref{soliton_dipole}. The vortex dipole is located at ($1$, 0) 
and ($-1$, 0) while the gray soliton extends from ($-1$, 0) to ($1$, 0) along 
the $x$-axis. In the case of vortex dipoles, the phase varies from 0 to $2\pi$, 
if one goes around the point of singularity. While in the case of gray soliton, 
there is a phase discontinuity of $\pi$ along the line forming the soliton. The 
number density at the point of singularities are zero.
In Fig. \ref{theoretical_pdf},  $E_s$ is plotted as a function of $v_2$ and 
the values varies from 0.05 $a_{\rm osc}$ to 2.0 $a_{\rm osc}$. From the 
figure it is evident that for $v_2\leqslant 0.2$, the value of $E_{\rm vd}$ is 
higher than $E_{\rm s}$ and hence, the gray soliton is the energetically 
favoured state of the system. However, when $v_2 > 0.2$ the vortex dipole state 
is the energetically favorable. This analytical result is a compelling reason 
to study the annihilation of vortex dipoles and formation of gray solitons.


\subsection{\label{subsec:gaussian}Weakly interacting system with gaussian 
approximation}

In $Na/a_{\rm osc}\ll1$ regime, a simplistic model of a vortex dipoles in the 
BEC of trapped dilute atomic gases can be considered as the superposition of 
harmonic oscillator eigenstates. The minimalist wave function which supports a 
vortex and antivortex at cordinates $(-a/c,-\sqrt{b/d})$ and 
$(-a/c,\sqrt{b/d})$ is 
\begin{equation}
  \psi(x, y) = e^{-i\mu t} \left (ia - b + i xc +dy^2 \right) 
    e^{-(x^2 + y^2)/f}, \label{vdip_sol}
\end{equation}
where $a$, $b$, $c$, $d$, and $f$ are positive variational parameters and $\mu$ 
is the chemical potential of the system. The 
wave function is a superposition of the scaled ground state and the first and 
the second excited states of harmonic oscillator along the $x$ and $y$-axes, 
respectively. The wave function is ideal for weakly interacting condensates.

\begin{figure}[h]
\begin{center}
   \includegraphics[width=6.5cm]{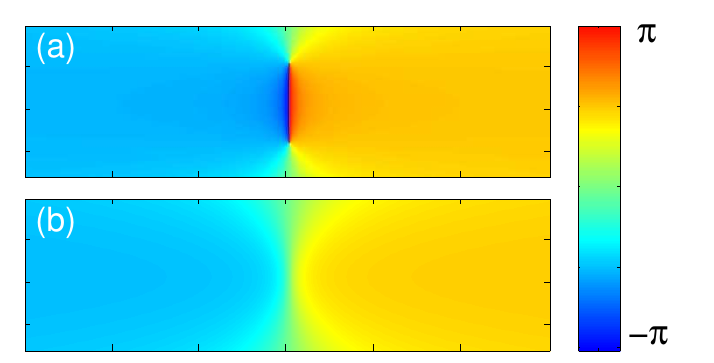}
   \caption{Phase pattern resulting due to a vortex-dipole (a) and 
   gray-soliton (b).}\label{phase_vd_vs_linear}
\end{center}
\end{figure}

  Consider that the vortex and antivortex are located on the diameter of the 
condensate. Without loss of generality, we consider the diameter as coinciding
with the $y$-axis, which is equivalent to $a=0$ in Eq.~\ref{vdip_sol}. Such an 
assumption does not modify the qualitative descriptions, but expressions are 
far less complicated. The wave function is then
\begin{equation}
   \psi(x,y,t) = e^{-i\mu t}\left[-b+icx+dy^2\right]e^{-(x^2 + y^2)/f}.
\end{equation}
The nontrivial phase of the wave function $\theta$ is discontinuous along 
$x=0$ line for $-\sqrt{b/d}\leqslant y\leqslant\sqrt{b/d}$. Across the 
discontinuity, there is a phase change from $-\pi$ to $\pi$ as we traverse 
along $x$-axis from $0^-$ to $0^+$ and this phase variation is shown in Fig. 
\ref{phase_vd_vs_linear}. So, there is a discontinuity across the $y$-axis
and this is the typical phase pattern associated with vortex dipoles. For the 
present case, the ground state wave function is 
\begin{equation}
    \psi_{\rm g}(x, y, t) = -b e^{-i\mu t} e ^{-(x^2 + y^2)/f},
\end{equation}
and from the normalization condition $\int_{-\infty}^\infty\int_{-\infty}^{ 
\infty}|\psi_{\rm g}|^2 \mathrm{d}x\mathrm{d}y = 1$, we get the constraint 
equation
\begin{equation}
   b^2 = \frac{2}{f\pi}.
  \label{norm_vort}
\end{equation}
For general considerations, rewrite the additional term as 
\begin{equation}
    \delta \psi(x, y, t) = e^{-i\mu t}\left (i cx +  dy^2 \right ) 
                           e ^{-(x^2 + y^2)/f}. 
\end{equation}
So that the total wave function $\psi=\psi_{\rm g}+\delta\psi$, where 
$\delta\psi$ represents an elementary excitation of the condensate. We can 
calculate the total energy of the system, without the obstacle potential, as 
\begin{eqnarray}
  E_{\rm vd} &=& \int_{-\infty}^{\infty}\int_{\infty}^{\infty}
                  \bigg [ \frac{1}{2}|\nabla_{\perp}\psi (x, y)|^2
                 + \frac{x^2+y^2}{2}|\psi(x, y)|^2 \nonumber \\
             &&  + u|\psi(x, y)|^4 \bigg ] \mathrm{d}x \mathrm{d}y .
\end{eqnarray}
This is the energy of the condensate with a vortex dipole with the assumption 
that it is a weakly interacting system. Energy without the vortex may be 
calculated trivially \cite{pethick}. In general, the energy added to the 
system due to the vortex dipole is not large compared to the total and for 
obvious reason, the angular momentum of the condensate is still zero.

A slight modification to the wave function can describe a solitonic solution
along $y$-axis. The form of the modifed wave function is
\begin{equation}
  \psi(x,y) = \left[b+i cx+dy^2\right]e^{-(x^2 + y^2)/f},\label{soli_sol} 
\end{equation}
where except for the change in the sign of $b$, all the terms remain unaltered
as in Eq. (\ref{vdip_sol}). It is a {\em gray} soliton as the density 
$n\propto (b+dy^2)^2+(a+cx)^2$ has a dip but is different from zero. The phase 
varies smoothly from $-\pi/2$ to $\pi/2$ along the normal to the line which 
connects $(0,-\sqrt{b/d})$ and $(0,\sqrt{b/d})$. This phase variation is
shown in Fig. \ref{phase_vd_vs_linear}(b).

Using the wave function in Eq.~(\ref{soli_sol}), we can then evaluate the total 
energy of the system $E_{\rm gs}$ and calculate the energy difference between
two possible states of the system
\begin{equation}
   \Delta E = E_{\rm vd} - E_{\rm gs},
\end{equation}
which after evaluation is 
\begin{equation}
   \Delta E = \frac{b d f^2\pi}{256} \left [ 64 b^2 u + 15 d^2 f^2 u 
             + 8 f (8 + c^2 u)\right].
\end{equation}
The most general solution is when all the constants are positive, then $\Delta 
E>0$ and the gray soliton is lower in energy. This shows that when the 
vortex-antivortex collides, it is energetically favourable for them to decay 
into gray soliton. As discussed in the results section, this is confirmed in 
the numerical calculations.

The analysis so far is for an ideal system at zero temperature, where we have 
neglected the thermal fluctuations and perturbations from imperfections. In 
addition, there is dissipation from three body collision losses in the 
condensates of dilute atomic gases.


\section{\label{sec:result}Numerical Results}

For the numerical computation,w e choose $^{87}$Rb with $N=2\times10^6$ atoms. 
The trapping potential and obstacle laser potential parameters are similar as 
those considered in Ref. \cite{PRL.104.160401}, i.e., $\omega/(2\pi)= 8$~Hz, 
$\beta=11.25$, $V_0(0)=93.0~\hbar\omega$ and $w_0=10~\mu$m. To nucleate the 
vortices on the edges of the condensate, the obstacle potential $V_{\rm obs}$ 
is initially located at $-12.5~a_{\rm osc}$ and moves along the $x$ direction 
at a constant velocity with decreasing intensity until $V_{\rm obs}$ vanishes 
at $5.18~a_{\rm osc}$.


\subsection{\label{subsec:nucleation}Vortex dipole nucleation}
We study the nucleation of vortices by $V_{\rm obs}$ with the translation speed 
$v$ ranging from $80~\mu$m/s to $200~\mu$m/s. Vortices are not nucleated when 
the speed is $80~\mu$m/s. However, a vortex-antivortex or a vortex dipole is 
nucleated when the speed is in the range $90~\mu$m/s $<v<140~\mu$m/s. 
Increasing the speed of obstacle generates two pairs of vortex dipoles 
when $140~\mu$m/s $\leqslant v<160~\mu$m/s and more than two when 
$v\geqslant 160~\mu$m/s. In other words, the number of vortex dipoles created 
can be controlled with the speed of the obstacle. Creation of vortex dipoles 
above a critical speed $v_c$ is natural as the vortex nucleation must satisfy 
the Landau criterion \cite{Lifshitz}. The density and phase of the condensate 
after the nucleation of vortex dipole for $v=120~\mu$m/s is shown in Fig. 
\ref{120_pdf}. The figure clearly shows nucleation dynamics of the vortex 
dipoles.

From numerical calculations, we have determined $v_c\approx90~\mu$m/s. 
This is, however, less than the local acoustic velocity of the medium 
$s=\sqrt{nU/m} $, which depends on the local condensate density. This also 
explains the reason for the predominant vortex dipole nucleation around the 
edge of the condensate where $n$ is lower and $s$ is accordingly lower.

\subsection{\label{subsec:annihilation}Vortex dipole annihilation}
\begin{figure}[h]
\begin{center}
   \includegraphics[width=8.3cm]{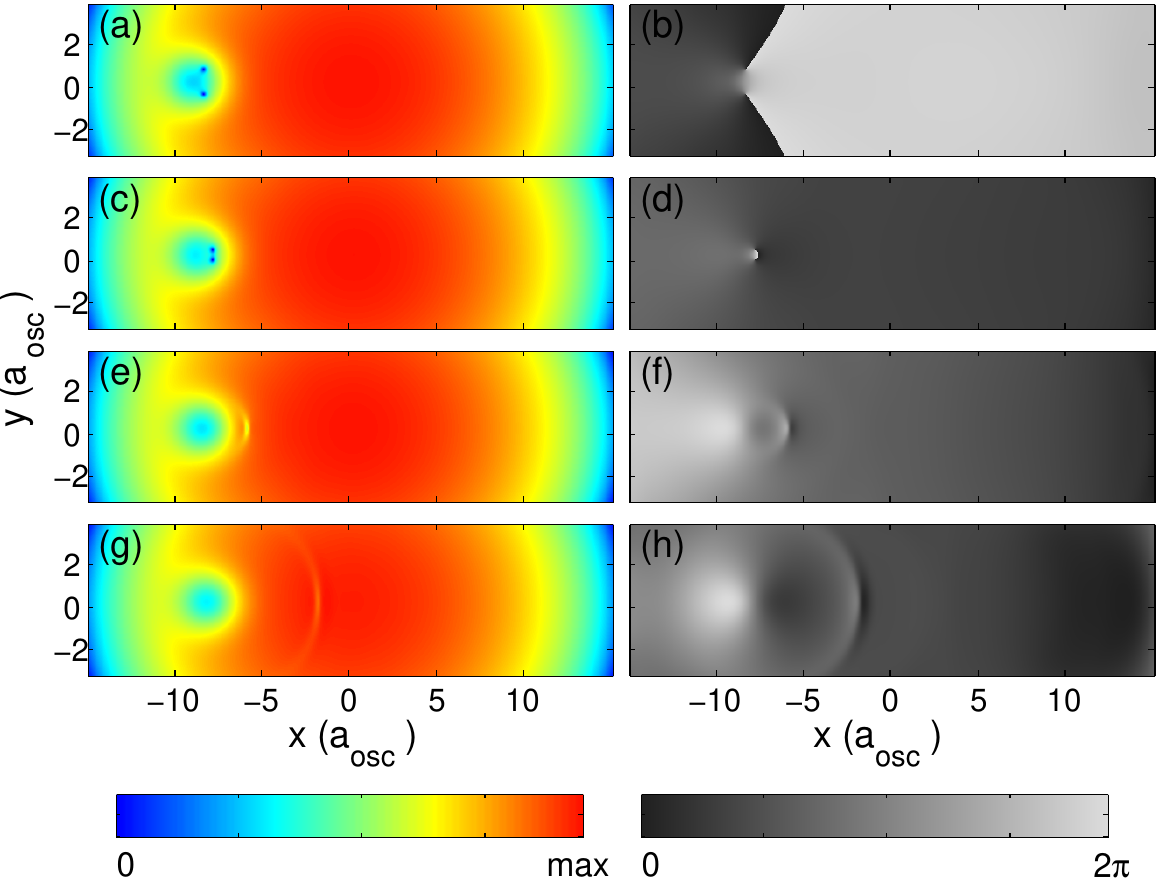}
   \caption{A vortex dipole is nucleated when the obstacle potential traverses 
     the condensate at a speed of $120~\mu$m/s. The vortex dipole, however, 
     passes through and overtakes the obstacle. Later, as seen in (e), the 
     vortex dipole annihilates and generates a gray soliton. The figures in the 
     left panel show the density distribution and those on the right show the 
     phase pattern of the condensate. From top to bottom, $t$ = 2.9, 3.1, 3.3, 
     and 3.5 respectively.}\label{120_pdf}
\end{center}
\end{figure}

It is observed that the vortex dipole annihilation is critically dependent
on the initial conditions of the nucleation. For this reason, the annihilation
events are observed only for specific range of $v$. As an example, the 
annihilation event when $v$ is $120~\mu$m/s is shown in Fig. \ref{120_pdf}.
In Fig. \ref{120_pdf}, we can notice the density minima arising from the 
annihilation and propagating away from the obstacle potential.

\begin{figure}[h]
\begin{center}
   \includegraphics[width=7.5cm]{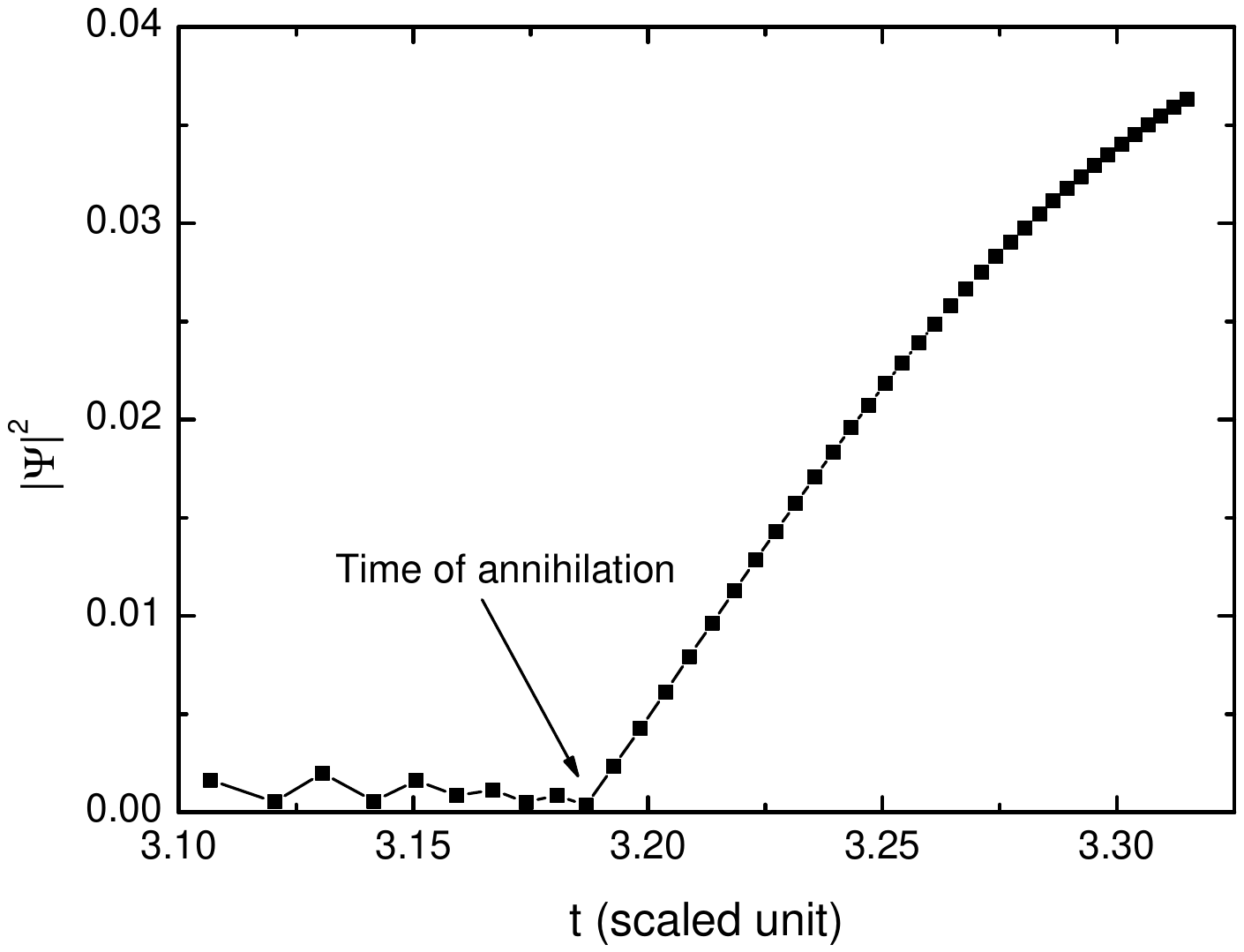}
   \caption{Density variation at the core of the vortex with time. After the 
     vortex dipole annihilation, density increases till it reaches the bulk 
     value. The values correspond to the obstacle speed of $120~\mu$m/s. After 
     annihilation, the number density has been considered from the location of 
     minimum density. X-axis denotes the time elapsed from the starting of 
     obstacle at $-12.5 a_{\rm osc}$.}\label{density_core}
\end{center}
\end{figure}

A reliable and qualitative way to describe occurrence of annihilation could be 
achieved by observing the density at the cores of vortex and antivortex which 
form the dipole. For the vortex, the matter density at the core when $v$ is 
$120~\mu$m/s is shown in Fig \ref{density_core}. In the plot, at time 
$\approx 3.1868$ (scaled unit), the core density starts increasing. This is 
because the core starts to fill in with the atoms from around the vortex after 
the annihilation. This filling process may not complete till it reaches the 
edge of the condensate and gets reflected inside the condensate.

\begin{figure}[h]
\begin{center}
   \includegraphics[width=6.5cm]{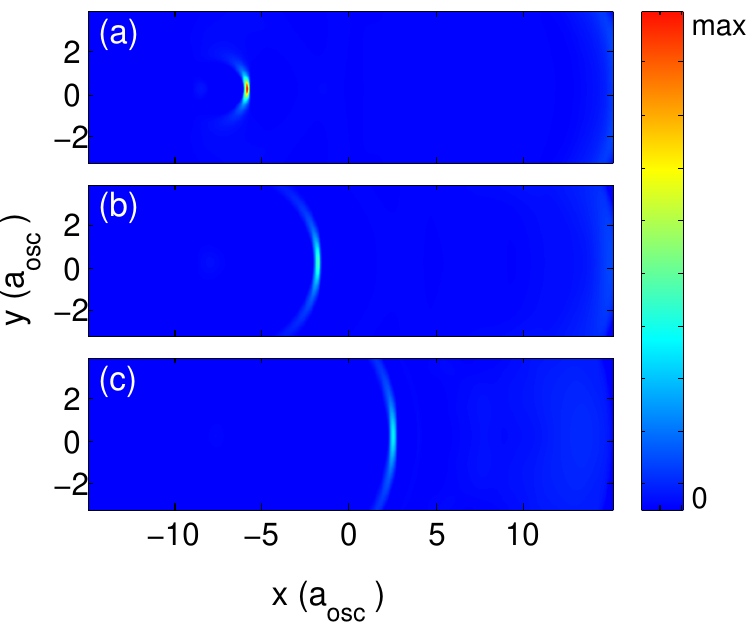}
   \caption{The propagation of the gray soliton after the annihilation of 
   vortex dipole. The higher the value, higher the number density dip at that 
   point. From (a)-(c), $t$ = 3.2, 3.4 and 3.6 respectively.}
   \label{sound_propagation}
\end{center}
\end{figure}

After the annihilation of vortex-antivortex dipole pair, a gray soliton gets 
generated. We can clearly observe the propagation of this soliton in Fig. 
\ref{sound_propagation}. The speed of propagation is 1999.6495 $\mu$m/s which 
is similar to the speed of sound in condensate. During the propagation, the 
number density on the location of the soliton increases which is clearly 
visible from Fig. \ref{sound_propagation} as well as from Fig. 
\ref{density_core}. To estimate the energy of gray soliton, we have obtained 
the stationary state with the same positon of vortex dipoles and obstacle 
potential. The energy difference between stationary state and dynamic state 
will provide us with the energy of gray soliton as discussed in ref 
\cite{adams}. The energy released due to the annihilation is 0.003995 
$\hbar\omega$ and is similar to the energy difference observed in Fig. 
\ref{theoretical_pdf}, obtained from the TF approximation. We have also 
observed that this soliton gets reflected back-and-forth from the edge of the 
condensate. This reflection is similar to the reflection of any pulse from 
the circular edges.

\begin{figure}[h]
\begin{center}
   \includegraphics[width=8cm]{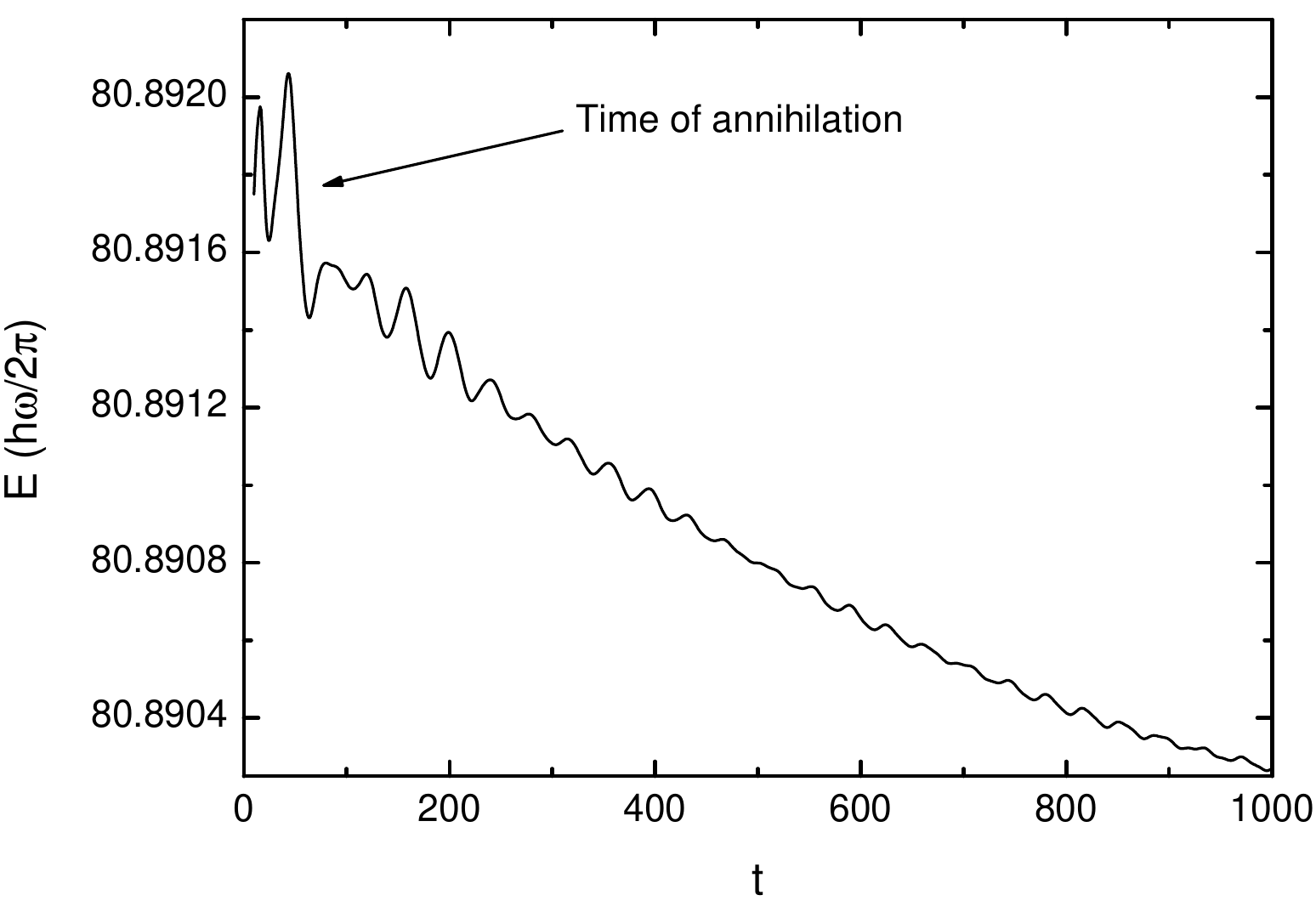}
   \caption{Variation in the energy of the system with time (in scaled units). 
     There is a decrease in energy after the vortex dipole annihilation, and 
     the plot is based on the results of calculations in which a vortex dipole 
     is imprinted at ($-2.0~a_{\rm osc}$, $\pm0.1~a_{\rm osc}$). It is then 
     allowed to annihilate and evolve in time.} \label{energy}
\end{center}
\end{figure}

As discussed in Section. \ref{sec:analytical}, annihilation can occur only when 
it is energetically favorable. In other words, the state with the gray soliton
must have lower energy than the vortex dipole. This is clearly seen in the 
change of total energy of the system, which is shown in Fig. \ref{energy}. 
After the annihilation, the energy of the system decreases and continues to
do so at a steady rate. Although not shown in the plot, before the annihilation 
the energy is on an average constant.

\begin{figure}[h]
\begin{center}   
   \includegraphics[width=8.3cm]{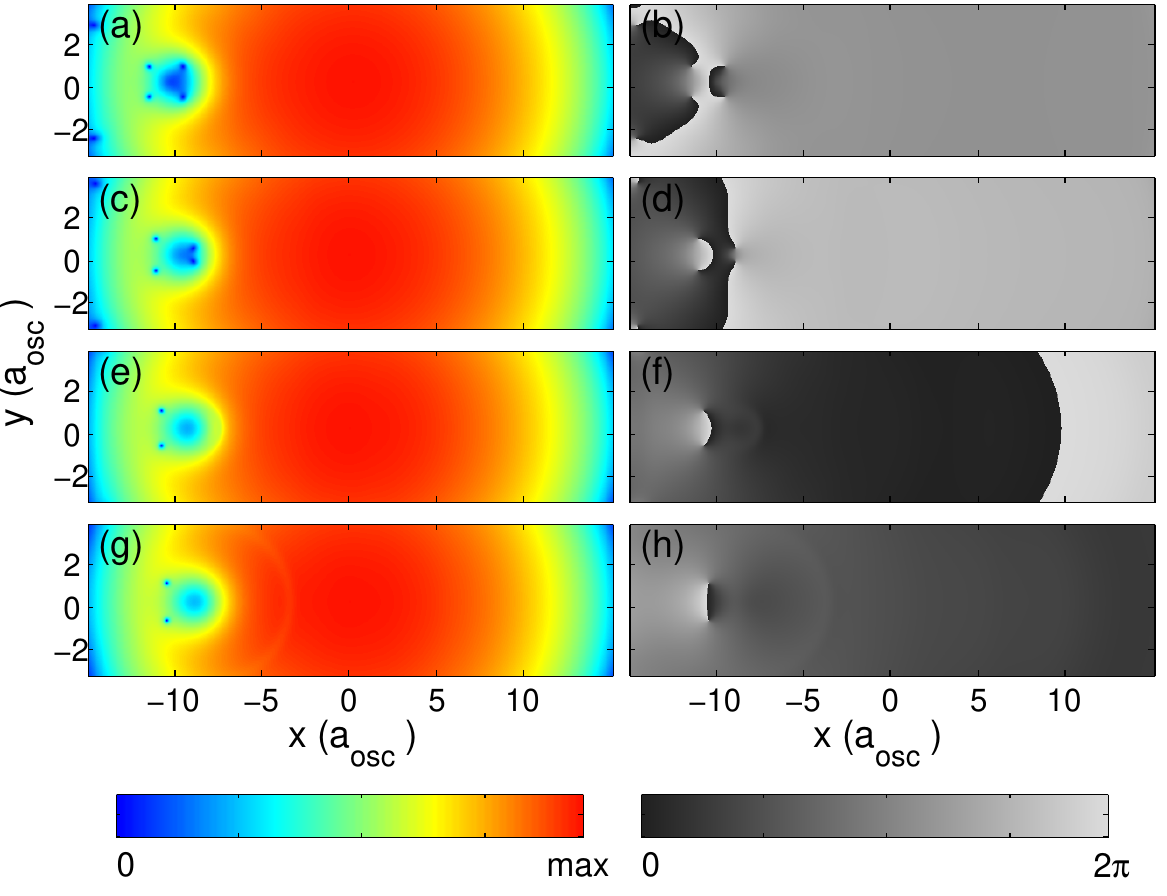}
   \caption{A vortex dipole is nucleated as the obstacle potential traverses 
     the BEC with a speed of $160~\mu$m/s. The figures in the left panel shows 
     the density with time, where time progresses from top to bottom. Figures 
     on the right panel show the phase pattern of the condensate. From top to 
     bottom, $t$ = 1.6, 1.8, 2.0, and 4.2 respectively.} \label{160_pdf}
\end{center}
\end{figure}

It is to be mentioned that for the parameters considered in the present work,
the speed of sound is $2190~\mu$m/s and the coherence length of the system is 
$\sim0.229~\mu$m. These are in agreement with the minimum separation between 
the vortex and antivortex observed in the analytical work. The energy gap for 
vortex dipole and gray soliton for the same size matches with the estimates 
from the ansatz based on TF approximation. The vortex dipole annihilation is 
not only observed for $V_{\rm obs} = 120~\mu$m/s, it also occures for other 
obstacle velocities as well. Once such case, for $V_{\rm obs} = 160~\mu$m/s, 
is shown in Fig. \ref{160_pdf}. In this case, the difference in energy of 
vortex dipole and gray soliton is 0.002481 $\hbar\omega$.

One observation, which is common to all the vortex dipoles getting annihilated 
is the nature of their trajectories. All of them traverse through $V_{\rm 
obs}$, whereas the ones which do not get annihilated avoid $V_{\rm obs}$. The 
vortex dipoles are generally nucleated at the aft region of the $V_{\rm obs}$ 
where there is a trailling superflow. When nucleated very close to each other 
and with high velocity, the mutual force further increases the velocity of the 
vortex dipoles. At the same time, it decreases the distance separating vortex 
and antivortex. So, the kinetic energy is high enough to surpass $V_{\rm obs}$. 
Later, at some point vortex and antivortex separation is less than $\xi$, and 
they annihilate.


\subsection{\label{subsec:noise}Effect of noise and dissipation}
\begin{figure}[h]
\begin{center}
   \includegraphics[width=8.3cm]{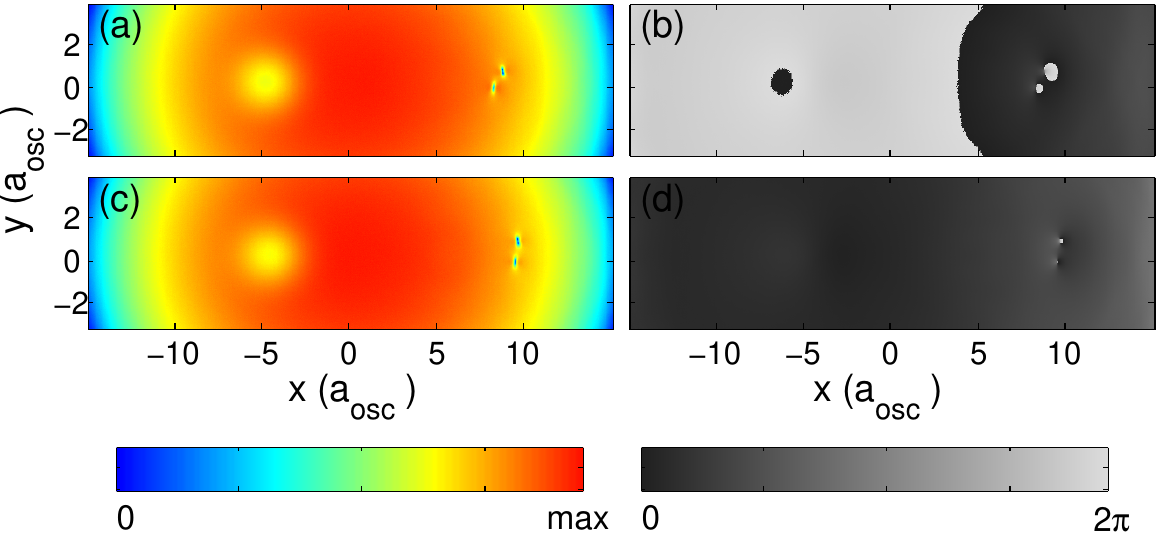}
   \caption{The trajectory of a vortex dipole in the presence of white noise.
     There is a lack of symmetry in the trajectory of the vortex and 
     antivortex. This reduces the possibility of an annihilation event 
     significantly. The figures in the left (right) panel show the density 
     (phase) of the condensate and time increases from top to bottom figures of 
     each panel. The speed of the obstacle is $180~\mu$m/s, and the white noise 
     is at the level of 0.01\%. Top and bottom correspond to $t = $ 4.1 and 4.2 
     respectively from the starting of obstacle at $x = -12.5 a_{\rm osc}$.}
     \label{noise}
\end{center}
\end{figure}

In the numerical studies, the annihilation events are not rare. But, this is in 
contradiction with the experimental results of Neely and collaborators 
\cite{PRL.104.160401}, they observed no signatures of annihilation events. One 
possible reason is that our numerical calculations are too ideal, and one 
immediate remedy is to include fluctuations. For this we introduce white noise 
during the real time evolution. One immediate outcome is, the symmetry in the 
trajectory of the vortex and antivortex is lost. The superflow around the 
vortex is no longer a mirror reflection of the antivortex, which was nearly the 
case without the white noise. The deviations are shown for an example case in 
the Fig. \ref{noise}, where $V_{\rm obs}=180~\mu$m/sec. The change in path 
leads to the suppression of annihilation of vortex dipoles.

The other important effect is the loss of atoms from the trap. We have examined
the effect of loss terms, which arise from inelastic collisions in the 
condensate. There are two types of inelastic collisions that lead to the loss 
of atoms from the trap: two body inelastic collision loss and the three body 
loss. To model the effect of loss of atoms from the trap, we add the loss terms 
\begin{equation}
   \frac{-i\hbar}{2} \left[ K_{2} |\Psi(\mathbf r,t)|^2 +K_3 |\Psi(\mathbf 
       r,t)|^4 \right],
\end{equation}
to the Hamiltonian ${\cal H}$. Based on the previous work \cite{PRL.80.2097} 
for $^{87}$Rb, the inelastic two-body loss rate coefficient $K_{2}=4.5\times 
10^{-17}~{\rm cm}^3~{\rm s}^{-1}$, and inelastic three-body loss rate 
coefficient $K_3=3.8\times10^{-29}~{\rm cm}^6~{\rm s}^{-1}$. With trap loss, 
the annihilation events continue to occur. However, during the destructive time 
of flight observations in the experiments, the decreased atom numbers may lower 
the contrast and reduce the possibility of observing an annihilation event.


\section{\label{sec:conclusions}Conclusions}
When an obstacle moves through a condensate above a critical speed, it 
nucleates vortex dipoles and the number of dipoles seeded depends on the 
obstacle velocity. Depending on the initial condition of nucleation, vortex and 
antivortex annihilation events occur under ideal conditions: at zero 
temperature, no loss, and without noise. These events are found to be 
thermodynamically favourable theoretically and observed numerically. In the 
case of weakly interacting condensates, the energy of gray soliton is always 
less than that of vortex dipole and provides higher possibility for 
annihilation events. Similary, in the case of strongly interacting condensates,
we use TF approximation to study the system and find that if the 
separation between the vortex anti-vortex pair is less then the coherence 
length, the energy of vortex dipole is more than that of gray soltion and this 
leads to annihilation. The gray soliton propagates through the condensate and 
shows the phenomena of reflection from the circular edge of the condensate. 
The speed of propagation is found to be similar to the speed of sound in BEC.
However, noise, thermal fluctuations and dissipation destroy superflow 
reflection symmetry around the vortex and antivortex. Breaking the symmetry 
reduces the possibility of annihilation events and may explain the lack of 
annihilation events in experimental observations.

\begin{acknowledgments}
The numerical calculations reported in this paper have been performed on 3 
TeraFlop high-performance cluster (HPC) at Physical Research Laboratory (PRL), 
Ahmedabad.
\end{acknowledgments}



\begin{thebibliography}{10}
\bibitem{Anglin2002}
J.~R. Anglin, and W.~Ketterle, Nature \textbf{416}, 211 (2002).

\bibitem{PRL.83.2498}
M.~R. Matthews, B.~P. Anderson, P.~C. Haljan, D.~S. Hall, C.~E. Wieman, and
  E.~A. Cornell, Phys. Rev. Lett. \textbf{83}, 2498 (1999).

\bibitem{PRL.104.160401}
T.~W. Neely, E.~C. Samson, A.~S. Bradley, M.~J. Davis, and B.~P. Anderson,
  Phys. Rev. Lett. \textbf{104}, 160401 (2010).

\bibitem{F03092010}
D.~V. Freilich, D.~M. Bianchi, A.~M. Kaufman, T.~K. Langin, and D.~S. Hall,
  Science \textbf{329}, 1182 (2010).
  
\bibitem{PRA.64.043601}
G.~Andrelczyk, M.~Brewczyk, \L{}.~Dobrek, M.~Gajda, and M.~Lewenstein, Phys. 
  Rev. A \textbf{64}, 043601 (2001).

\bibitem{PRA.65.043612}
J.~Brand, and W.~P. Reinhardt, Phys. Rev. A \textbf{65}, 043612 (2002).

\bibitem{PRA.84}
S.~Middelkamp, P.~J.~Torres, P.~G.~Kevrekidis, D.~J.~Frantzeskakis, 
  R.~Carretero-Gonz\'alez, P.~Schmelcher, D.~V.~Freilich, and D.~S.~Hall, Phys. 
  Rev. A \textbf{84}, 011605(R) (2011).

\bibitem{PRX.1}
T.~Aioi, T.~Kadokura, T.~Kishimoto, and H.~Saito, Phys. Rev. X \textbf{1}, 
  021003 (2011).

\bibitem{CJ:392656}
Y.~Couder, and C.~Basdevant, J. Fluid. Mech. \textbf{173}, 225 (1986).

\bibitem{Grier2003}
D.~G. Grier, Nature \textbf{424}, 810 (2003).

\bibitem{pethick}
C.~Pethick, and H.~Smith, \emph{Bose-Einstein Condensation in Dilute Gases}
  (Cambridge University Press, 2002).

\bibitem{rmp.81.647}
A.~L. Fetter, Rev. Mod. Phys. \textbf{81}, 647 (2009).

\bibitem{rmp.59.87}
E.~B. Sonin, Rev. Mod. Phys. \textbf{59}, 87 (1987).

\bibitem{okulov}
S.~Alekseenko, P.~Kuibin, and V.~Okulov, \emph{Theory of Concentrated Vortices}
  (Cambridge University Press, 2002).

\bibitem{PRA.61.013604}
B.~Jackson, J.~F. McCann, and C.~S. Adams, Phys. Rev. A \textbf{61}, 013604
  (1999).

\bibitem{PRA.77.053610}
W.~Li, M.~Haque, and S.~Komineas, Phys. Rev. A \textbf{77}, 053610
  (2008).

\bibitem{JLTP.146.31}
S.~Nazarenko, and M.~Onorato, J. Low. Temp. Phys. \textbf{146}, 146 (2007).

\bibitem{PRA.84.023637}
S.~J.~Rooney, P.~B.~Blakie, B.~P.~Anderson, and A.~S.~Bradley, Phys. Rev. A
  \textbf{84}, 023637 (2011).

\bibitem{PRB.84.020506}
B.~Nowak, D.~Sexty, and T. Gasenzer, Phys. Rev. B \textbf{84}, 020506(R) (2011).

\bibitem{prl88}
E.~G.~M.~van Kempen, S.~J.~J.~M.~F.~Kokkelmans, D.~J.~Heinzen, and 
  B.~J.~Verhaar, Phys. Rev. Lett. \textbf{88}, 093201 (2002).

\bibitem{Muruganandam20091888}
P.~Muruganandam, and S.~Adhikari, Comput. Phys. Commun. \textbf{180},
  1888 (2009).

\bibitem{PRL.80.2972}
R.~Dum, J.~I. Cirac, M.~Lewenstein, and P.~Zoller, Phys. Rev. Lett.
  \textbf{80}, 2972 (1998).

\bibitem{PRL.79.4728}
K.~P. Marzlin, W.~Zhang, and E.~M. Wright, Phys. Rev. Lett. \textbf{79}, 4728
  (1997).

\bibitem{s003400000337}
K.~Staliunas, Applied Physics B: Lasers and Optics \textbf{71}, 555 (2000).

\bibitem{PRA.83.011603}
P.~Kuopanportti, J.~A.~M. Huhtam\"aki, and M.~M\"ott\"onen, Phys. Rev. A
  \textbf{83}, 011603 (2011).

\bibitem{Zhou}
Q.~Zhou, and H.~Zhai, Phys. Rev. A \textbf{70}, 043619 (2004).

\bibitem{Lifshitz}
E.~M.~Lifshitz, and L.~P.~Pitaevskii, \emph{Statistical Physics, Part 2} 
  (Pergamon Press, Oxford, 1980).

\bibitem{adams}
N.~G.~Parker, and C.~S.~Adams, Phys. Rev. Lett. \textbf{95}, 145301 (2005).

\bibitem{PRL.80.2097}
J.~P.~Burke, J.~L.~Bohn, E.~D.~Esry, and C.~H.~Greene, Phy. Rev. Lett. 
  \textbf{80}, 2097 (1998).

\end{thebibliography}
\end{document}